# Magnetic skyrmion logic gates: conversion, duplication and merging of skyrmions

Xichao Zhang[1], Motohiko Ezawa[2, *], Yan Zhou[1, 3, †]

[1]Department of Physics, University of Hong Kong, Hong Kong, China, [2]Department of Applied Physics, University of Tokyo, Hongo 7-3-1, Tokyo 113-8656, Japan, [3]Center of Theoretical and Computational Physics, University of Hong Kong, Hong Kong, China

Magnetic skyrmions, which are topological particle-like excitations in ferromagnets, have attracted a lot of attention recently. Skyrmionics is an attempt to use magnetic skyrmions as information carriers in next generation spintronic devices. Proposals of manipulations and operations of skyrmions are highly desired. Here, we show that the conversion, duplication and merging of isolated skyrmions with different chirality and topology are possible all in one system. We also demonstrate the conversion of a skyrmion into another form of a skyrmion, *i.e.*, a bimeron. We design spin logic gates such as the AND and OR gates based on manipulations of skyrmions. These results provide important guidelines for utilizing the topology of nanoscale spin textures as information carriers in novel magnetic sensors and spin logic devices.

Correspondence and requests for materials should be addressed to:
[*]M.E. (ezawa@ap.t.u-tokyo.ac.jp)
[†]Y.Z. (yanzhou@hku.hk)

Skyrmions are topologically stable field configurations with particle-like properties originally introduced in the context of particle physics and later extended to various fields of science [1-21]. Most recently, isolated skyrmion has been experimentally demonstrated in magnetic thin film with Dzyaloshinskii–Moriya interaction (DMI) [22]. Experimental realizations of skyrmions have excited a flourish of study of this nanoscale magnetic texture [23-32]. Topological stability of a skyrmion makes it promising for future applications in non-volatile memory and spintronics devices [17, 18, 27-29, 33].

There are many types of magnetic quasi-particles such as domain walls, skyrmions and merons, which will be useful for the potential applications of ultra-dense information storage and logic devices. To explore these intriguing magnetic nano-objects for multifunctional spintronic applications, it is crucial to realize their mutual conversions and transmissions in order to fully utilize their combined advantages in circuits and devices based on these magnetic excitations. In this work, we demonstrate the conversion, duplication, merging and collapse of topologically rich spin-vortex configurations including skyrmions and merons all in one system through spin transfer torque. Furthermore, we show that spin logic gate such as the AND and OR gates can be constructed by using skyrmions, which opens a new area by utilizing topological skyrmions to carry digital information in nanowire junctions. Our study will have significant impact on a range of emerging spintronic applications by adding an entirely new dimension to the rapid progressing field of skyrmionics and may bring about theoretical breakthrough in terms of fundamental topological field theory and concepts.

A skyrmion is topologically stable because it carries a quantized topological number [1, 2]. This is absolutely correct in particle physics [3, 4]. However, in condensed matter physics, there are exceptional cases. Our basic observation is that the skyrmion number can be changed in two ways: (i) A skyrmion can be destroyed or created at the edge which modified the spin direction of the tail of a skyrmion [24, 27]. (ii) It can be destroyed or generated by breaking the continuity of spin. It has been demonstrated that a skyrmion can be generated from a notch [24] or by converting a domain-wall pair at a narrow-wide junction [27]. By breaking the continuity of spin, a skyrmion can be generated by photo-irradiation [25, 26] or destroyed by shrinking its size to the scale of the lattice constant.

Our setup is shown in Fig. 1a, where the left input and right output wide regions are connected by a narrow nanowire. This is a composite structure made of two narrow-wide nanowire junctions. We may assume that the left input and right output regions have different material parameters, as shown in Fig. 1b, in the left input side the magnetization is pointing up, and the DMI is positive. We may assume the DMI in the right output side is negative in Fig. 2b, and the magnetization in the right output side is pointing down in Fig. 2c. Both the DMI and the magnetization in the right output side are opposite in Fig. 2d as compared to Fig. 2a. A smooth gradient transition of parameter from the value of input side to the value of output side is used in the narrow nanowire.

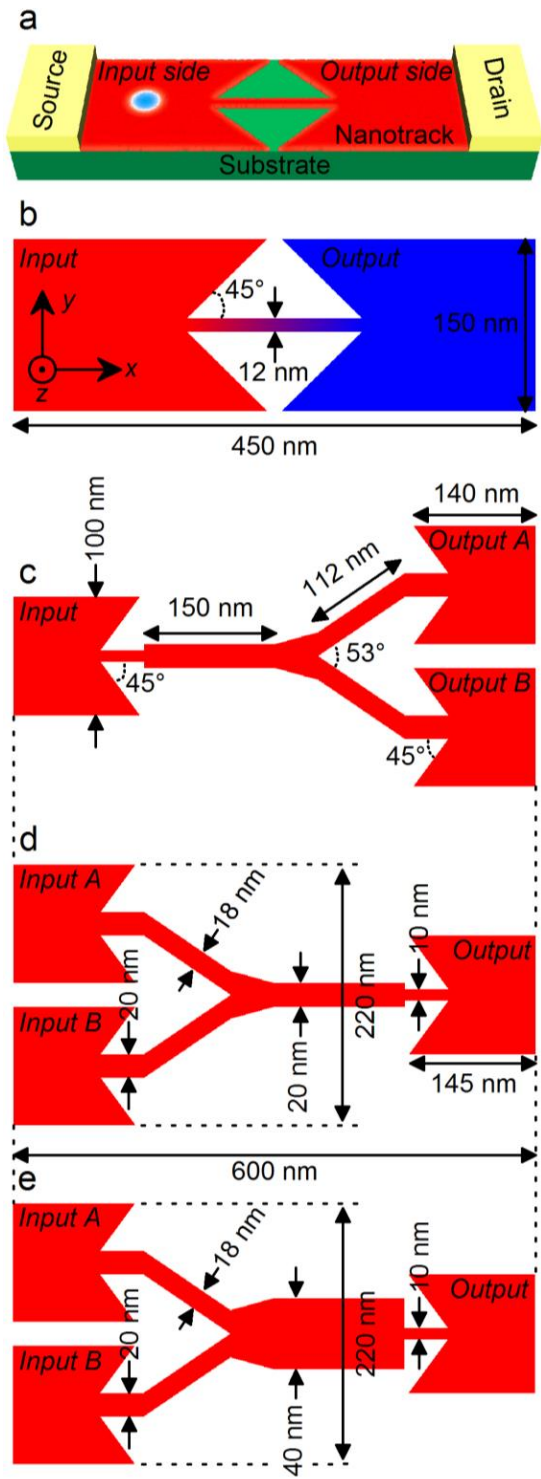

**Figure 1. The basic design of the magnetic skyrmion logic gate system. a**, Sketch of the simulated model: the red and green layers represent the nanowire and the substrate, respectively; the spin-polarized current is injected into the nanowire with the current-in-plane (CIP) geometry, through which electrons flow from the source to the drain, *i.e.*, toward +*x*; the current density inside the wide part of the nanotrack is proportional to the current density inside the narrow part of the nanotrack with respect to the ratio of narrow width to wide width; a skyrmion is initially created at the input side and can be pushed into the output side by the current with the conversion between skyrmion and domain-wall pair in the junction geometry. **b**, The top-view of the design of the 1-nm-thick skyrmion-conversion geometry: the width of the input and output sides is 150 nm, the width of the narrow channel is 12 nm, and the length of the sample is 450 nm; the interface connection angle is fixed at 45 degrees (similarly hereinafter); red and blue denote two regions with different parameters, where a gradient transition of parameter is used on the narrow channel. **c**, The top-view of the design of the 1-nm-thick geometry for the skyrmion duplication: the width of all the input and output sides is 100 nm, and the length of the sample is 600 nm. **d**, The top-view of the design of the 1-nm-thick geometry for the skyrmion merging and the logical OR gate: the geometry is the horizontally-flipped version of the one in **c**. **e**, The top-view of the design of the 1-nm-thick geometry for the logical AND gate: the geometry is similar to the one in **d**, except the horizontal branch of the Y-junction channel, of which the width is increased from 20 nm to 40 nm. The current density inside the output side is equal to the sum of that inside the two input sides. All the designed samples can connect to nanowires with matching width of the branch for application in integrated circuit devices.

A skyrmion is characterized by three numbers: the Pontryagin number $Q_s$, the vorticity $Q_v$ and the helicity $Q_h$. It is called a skyrmion (anti-skyrmion) when this Pontryagin number $Q_s$ is positive (negative). Vorticity of a skyrmion is defined by the winding number of the spin configurations projected into the $s_x$-$s_y$ plane. As is shown in Method, the skyrmion number is determined by two properties; the spin direction at the core and the tail of a skyrmion, and the vorticity. We will show that it is possible to convert a skyrmion into an anti-skyrmion or vice versa by changing the spin direction at the core and the tail (see Fig. 2c). On the other hand, the helicity does not contribute to the topological number. It is uniquely determined by the type of the DMI. A skyrmion with the helicity 0 and $\pi$ corresponds to the Néel-type skyrmion, while a skyrmion with the helicity $\pi/2$ and $3\pi/2$ corresponds to the Bloch-type skyrmion. We will show that the Néel-type skyrmion with the helicity 0 can be transformed into a skyrmion with the helicity of $\pi$ (see Fig. 2b).

Skyrmionics refers to the attempt to use skyrmions for device applications. It remains still a primitive stage and the operations of skyrmions such as logic gates are lacking. Furthermore, the duplication of information is an important process of skyrmionics, which has not been addressed yet. The conversion of skyrmions in different media is also important for realization of skyrmionic devices. In this paper, we propose a unified system where all these processes can be realized.

## Results

**Skyrmion conversion between different magnetic materials.** First we investigate what happens when the material parameters are uniform. The result is shown in Fig. 2a (see Supplementary Movie S1). A skyrmion in the left input wide nanowire moves rightward by the injected current, which is converted into a domain-wall pair. This is the conversion between a domain-wall pair and a skyrmion, which is reported previously [27]. In the central narrow nanowire region, the domain-wall pair continues to move rightward. When the domain-wall pair reaches the junction, it is converted into a skyrmion, which is an inverse process of the domain wall-skyrmion (DW-skyrmion) conversion. In this process the quantum numbers ($Q_s$, $Q_v$, $Q_h$) change as (1, 1, 0) → (0, 0, 0) → (1, 1, 0).

Interesting situations occur when the sign of the DMI is opposite between the left and right samples. Recent experiment has shown that one can experimentally control the magnitude and the sign of the DMI by changing continuously the component of materials [34]. Thus our setup is experimentally feasible. Fig. 2b shows the conversion of a skyrmion with different helicity via a domain-wall pair (see Supplementary Movie S2). We have assumed the sign of the DMI is positive in the left input region, while it is negative in the right output region. A gradient transition of the sign of the DMI from positive to negative is set in the narrow channel. The helicity is uniquely determined by the DMI. Accordingly, a skyrmion has the helicity of 0 in the left region, while it has the helicity of π in the right region. Namely, the spin direction of the skyrmion in the left input region is out-going, while that in the right output region is in-going. In this process the quantum numbers change as (1, 1, 0) → (0, 0, 0) → (1, 1, π).

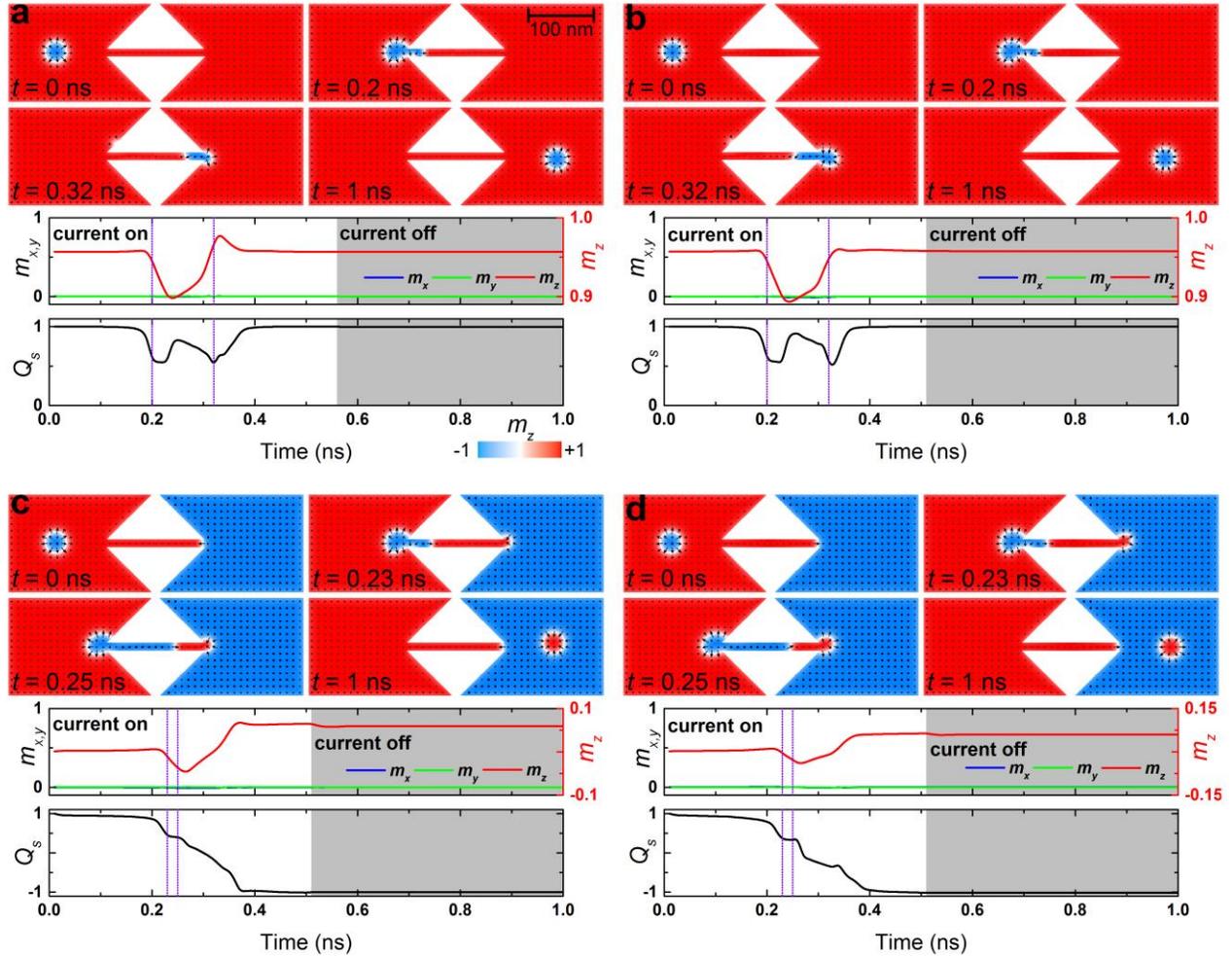

**Figure 2. Conversions between skyrmions and antiskyrmions.** The top panels show the snapshots of the magnetization configuration at four selected times corresponding to the vertical lines in the middle and bottom panels; the middle panels show the time evolution of the average spin components $m_x$, $m_y$, $m_z$; the bottom panels show the time evolution of the skyrmion number $Q_s$. **a**, Conversion between a skyrmion and a skyrmion with identical out-going helicity: the $D$ in the sample is 3.5 mJ m$^{-2}$; the background points +$z$; a current density of 3×10$^{12}$ A m$^{-2}$ (the value is of the input or output side, similarly hereinafter) is applied along -$x$ for 0 ns < $t$ < 0.56 ns followed by a relaxation (highlighted by the gray shadows) until $t$ = 1 ns. **b**, Conversion between a skyrmion and a skyrmion with opposite in-going helicity: the $D$ is 3.5 mJ m$^{-2}$ in the input side and -3.5 mJ m$^{-2}$ in the output side, while a gradient transition from 3.5 mJ m$^{-2}$ to -3.5 mJ m$^{-2}$ is set in the narrow channel; the background points +$z$; a current density of 3×10$^{12}$ A m$^{-2}$ is applied along -$x$ for 0 ns < $t$ < 0.51 ns and then is the relaxation until $t$ = 1 ns. **c**, Conversion between a skyrmion and an anti-skyrmion with opposite in-going helicity: the $D$ in the sample is 3.5 mJ m$^{-2}$; the background of the input side points +$z$, while it points -$z$ in the output side; a current density of 2.67×10$^{12}$ A m$^{-2}$ is applied along -$x$ for 0 ns < $t$ < 0.51

ns followed by a relaxation until $t$ = 1 ns. **d**, Conversion between a skyrmion and an anti-skyrmion with identical out-going helicity: the profile of *D* is the same as that in **b** and the profile of background is the same as that in **c**; a current density of 2.67×10$^{12}$ A m$^{-2}$ is applied along *-x* for 0 ns < $t$ < 0.51 ns followed by a relaxation until $t$ = 1 ns. The color scale has been used throughout this paper.

Next we investigate the system where the magnetization between the left and right regions is opposite without changing the DMI. It can be realized, *e.g.*, by applying an external magnetic field with opposite direction in the left input and the right output regions. As is shown in the Method section, the topological number is determined by the difference of the spin direction at the core and the tail. The spin direction of the tail must be the same as the background magnetization. Accordingly, the stable structure in the left input region is a skyrmion, while in the right output region it is an anti-skyrmion. By connecting the left and right samples via a narrow nanowire, we can convert a skyrmion into an anti-skyrmion, as shown in Fig. 2c (see Supplementary Movie S3). In the conversion process between a skyrmion and an anti-skyrmion, the helicity is also reversed to minimize the energy (see Method for details). In this process the quantum numbers change as (1, 1, 0) → (0, 0, 0) → (-1, 1, π).

Figure 2d shows the result when both the sign of the DMI and the direction of the magnetization are opposite between the left input and right output regions (see Supplementary Movie S4). In this case, the skyrmion with the helicity of 0 is converted into an anti-skyrmion with the helicity of 0. In this process the quantum numbers change as follows: (1, 1, 0) → (0, 0, 0) → (-1, 1, 0).

Another important material parameter is the magnetic anisotropy, which determines whether the spins favor easy-axis or easy-plane anisotropy. So far we have assumed the easy-axis anisotropy, where a skyrmion is stable. However, a sample with easy-plane anisotropy is also interesting. In this case, a skyrmion cannot take its standard form since the spin direction of the tail should be in plane. For the conventional skyrmions in materials with perpendicular magnetic anisotropy (PMA), the tail must be up or down direction, which topologically protects the stability. In the easy-plane sample, the tail must be in-plane. A vortex and an anti-vortex may be a simple realization of this boundary condition. However they cannot exist in ferromagnet in a strict sense since they do not preserve the norm of spin. Namely, at the core of a vortex and anti-vortex, the norm of the spin must be zero. On the other hand, it is known that a meron and anti-meron can exist in ferromagnetic materials. The tail of a meron (anti-meron) is exactly the same of that of a vortex (anti-vortex). However the spin direction of the core of a meron and an anti-meron is up or down, which preserves the norm of spin. As a result, a meron (anti-meron) has a skyrmion number of 1/2 (-1/2). In this sense a meron (anti-meron) is a "half-skyrmion". However it cannot exist by itself in a sufficiently large sample since the energy of tail diverges at the infinity. This is exactly the same as the case of a vortex. The stable form which has a finite energy is a bimeron, which is a pair of merons. The energy of a bimeron does not diverge since the tail far away from the core is parallel to the spin direction of the ground state. A bimeron (anti-bimeron) has a skyrmion number of 1 (-1). In this sense a bimeron (anti-bimeron) is another form of a skyrmion which exists in easy-plane sample. The results are shown in Fig. 3a and 3b, where a skyrmion is converted into a bimeron (anti-bimeron) for two cases of the stable initial magnetization state of the easy-plane sample (see Supplementary Movies S5 and S6). We note that there is a rotational degree of freedom for easy-plane anisotropy. The initial spin direction of the right region is pointing leftward in Fig.3a, while that is pointing rightward in Fig.3b. An anti-bimeron with (-1, 0, 0) is formed in the case of Fig. 3a and a bimeron with (1, 0, 0) is formed in the case of Fig. 3b. These results show that the conversion mechanism is independent of the initial spin direction of the right region.

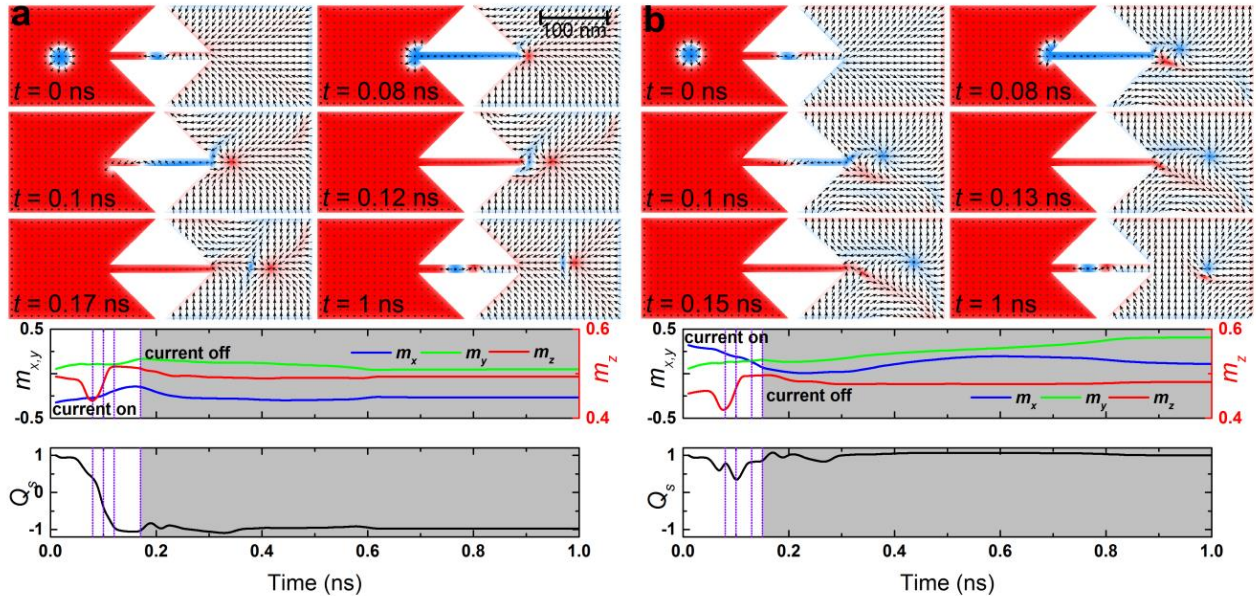

**Figure 3. Conversion between a skyrmion and a bimeron.** The top panels show the snapshots of the magnetization configuration at six selected times corresponding to the vertical lines in the middle and bottom panels; the middle panels show the time evolution of the average spin components $m_x$, $m_y$, $m_z$; the bottom panels show the time evolution of the skyrmion number $Q_s$. **a**, Conversion between a skyrmion and an anti-bimeron: the $D$ in the sample is 3.5 mJ m$^{-2}$; the anisotropy $K$ is 0.8 MJ m$^{-3}$ in the input side and -0.8 MJ m$^{-3}$ in the output side, while a gradient transition from 0.8 MJ m$^{-3}$ to -0.8 MJ m$^{-3}$ is set in the narrow channel, *i.e.* the plane of the input side is a hard plane, while the plane of the output side is an easy plane. The initial background magnetization of the input side points +z, while it is mostly aligned along -x direction in the output side; a current density of 9×10$^{12}$ A m$^{-2}$ (the value is of the input or output side, similarly hereinafter) is applied along -x direction for 0 ns < $t$ < 0.17 ns followed by a relaxation without applying any current until $t$ = 1 ns. **b**, Conversion between a skyrmion and a bimeron: the $D$ is 3.5 mJ m$^{-2}$; the profile of the anisotropy is the same as that in **a**. The initial background magnetization of the input side points +z, while it is mostly aligned along +x direction in the right output side; a current density of 10×10$^{12}$ A m$^{-2}$ is applied along -x direction for 0 ns < $t$ < 0.15 ns followed by a relaxation until $t$ = 1 ns.

**Duplication and merging of skyrmions.** For potential applications of devices based on skyrmions, it is important to copy information in addition to transport information. This can be accomplished by the duplication process of a skyrmion, which is shown in Fig. 4a (see Supplementary Movie S7). The setup is shown in Fig. 1c. It should be noted that as the opening angle of the Y-junction could be important for spintronic application [35], we also studied serval setups with different angle of the Y-junction (See Supplementary Information). A skyrmion is converted into a domain-wall pair. At the Y-junction in the central region, the domain-wall pair is split into two domain-wall pairs. Then they are converted into two skyrmions. As a result, one skyrmion becomes two skyrmions, which is a duplication of a single skyrmion. The information is stored by the position and timing of a skyrmion. Then the information is also duplicated using this system. In this process the quantum numbers change as (1, 1, 0) → (0, 0, 0) → (2, 1, 0).

An inverse process is also possible. Namely, two skyrmions merge into one skyrmion via a domain-wall pair as shown in Fig. 4b (see Supplementary Movie S8). In this process the quantum numbers change as (2, 1, 0) → (0, 0, 0) → (1, 1, 0).

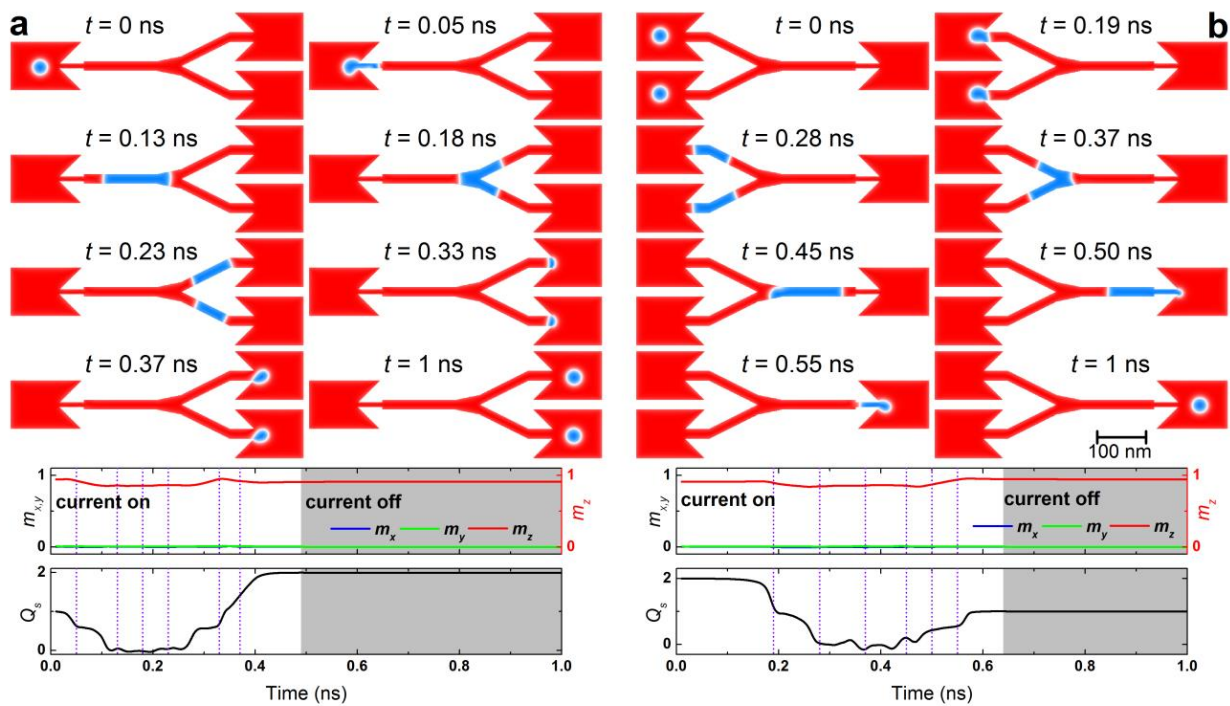

**Figure 4. Duplication and merging of skyrmion.** The top panels show the snapshots of the magnetization configuration at eight selected times corresponding to the vertical lines in the middle and bottom panels; the middle panels show the time evolution of the average spin components $m_x$, $m_y$, $m_z$; the bottom panels show the time evolution of the skyrmion number $Q_s$. **a**, Duplication of a skyrmion: the $D$ is 3.5 mJ m$^{-2}$; the initial background magnetization of the sample points $+z$; a current density of 5×10$^{12}$ A m$^{-2}$ (the value is of the input

side) is applied along -*x* direction for 0 ns < *t* < 0.49 ns followed by a relaxation without applying any current until *t* = 1 ns. **b**, Merging of two skyrmions: the *D* is 3.5 mJ m$^{-2}$; the initial background magnetization of the sample points +*z*; a current density of 4×10$^{12}$ A m$^{-2}$ (the value is of the output side) is applied along -*x* direction for 0 ns < *t* < 0.64 ns followed by a relaxation until *t* = 1 ns.

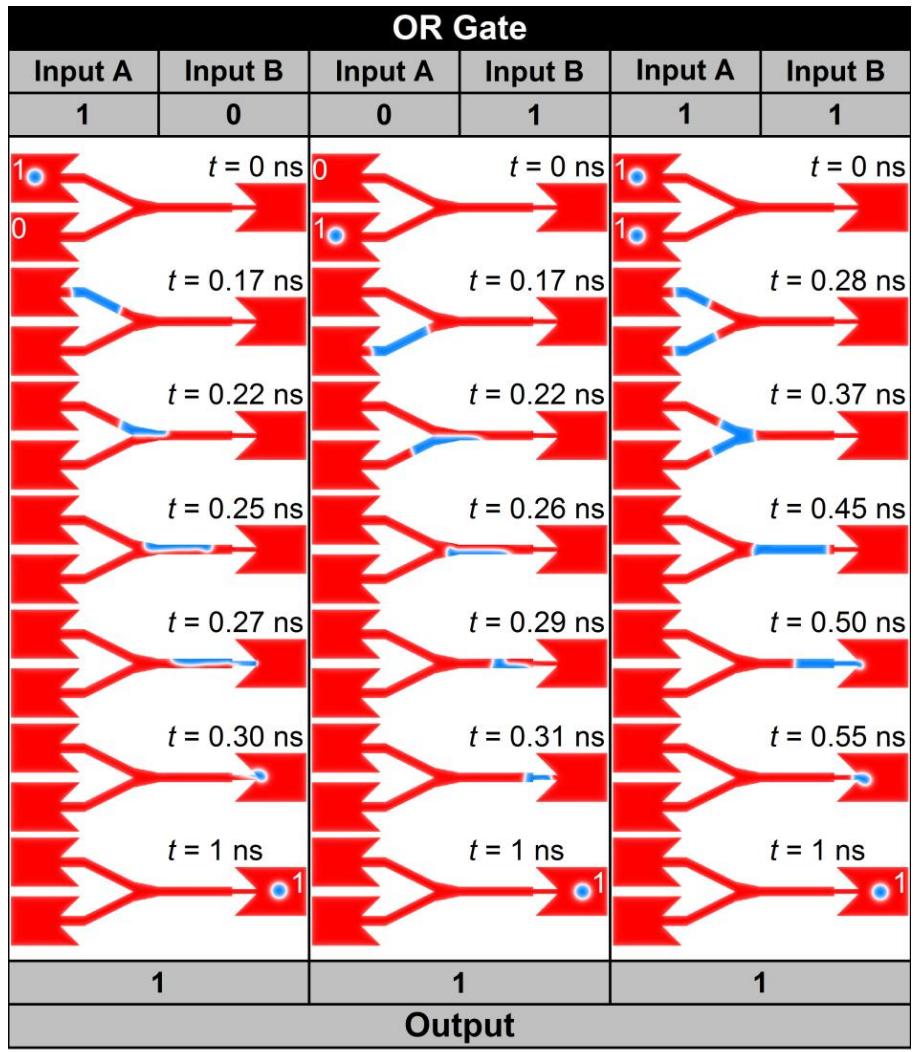

**Figure 5. Skyrmion logical OR operation.** The skyrmion represents logical 1, and the ferromagnetic ground state represents logical 0. **Left panel**, the basic operation of OR gate 1 + 0 = 1: there is a skyrmion in the input A and no skyrmion in the input B at initial time, which represents input = 1 + 0; a current density of 7×10$^{12}$ A m$^{-2}$ (the value is of the output side, similarly hereinafter) is applied along -*x* direction for 0 ns < *t* < 0.39 ns followed by a relaxation without applying any current until *t* = 1 ns. At *t* = 1 ns, a stable skyrmion is in the output side, which represents output = 1. **Middle panel**, the basic operation of the OR gate 0 + 1 = 1: there is a skyrmion in the input B side and no skyrmion in the input A side at initial time, which represents input = 0 + 1; a current density of 7×10$^{12}$ A m$^{-2}$ is applied along -*x* direction for 0 ns < *t* < 0.39 ns followed by a relaxation without applying any current until *t* = 1 ns. At *t* = 1 ns, a stable skyrmion is in the output side, which represents output = 1. **Right panel**, the basic operation of the OR gate 1 + 1 = 1: there is a skyrmion in both the input A side and the input B side, which represents input = 1 + 1; a current density of 4×10$^{12}$ A m$^{-2}$ is applied along -*x* direction for 0 ns < *t* < 0.64 ns followed by a relaxation without applying any current until *t* = 1 ns. At *t* = 1 ns, a stable skyrmion is in the output side, which represents output = 1.

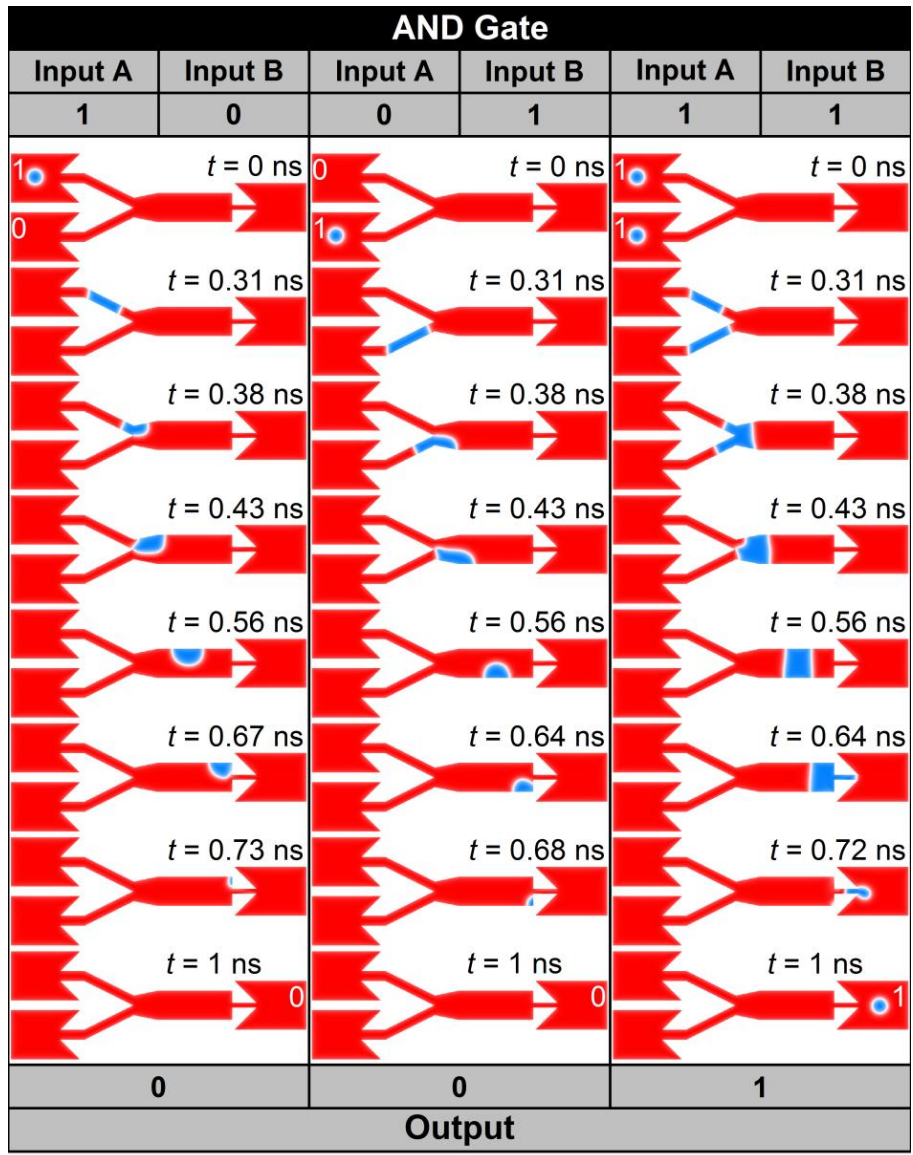

**Figure 6. Skyrmion logical AND operation.** The skyrmion represents logical 1, and the ferromagnetic ground state represents logical 0. **Left panel**, the basic operation of AND gate 1 + 0 = 0: there is a skyrmion in the input A side and no skyrmion in the input B side at initial time, which represents input = 1 + 0; a current density of $4\times10^{12}$ A m$^{-2}$ (the value is of the output side, similarly hereinafter) is applied along -$x$ direction for 0 ns < $t$ < 0.81 ns followed by a relaxation without applying any current until $t$ = 1 ns. At $t$ = 1 ns, no skyrmion is in the output side, which represents output = 0. **Middle panel**, the basic operation of the AND gate 0 + 1 = 0: there is a skyrmion in the input B side and no skyrmion in the input A side at initial time, which represents input = 0 + 1; a current density of $4\times10^{12}$ A m$^{-2}$ is applied along -$x$ direction for 0 ns < $t$ < 0.81 ns followed by a relaxation without applying any current until $t$ = 1 ns. At $t$ = 1 ns, no skyrmion is in the output side, which represents output = 0. **Right panel**, the basic operation of the AND gate 1 + 1 = 1: there is a skyrmion in both the input A side and the input B side, which represents input = 1 + 1; a current density of $4\times10^{12}$ A m$^{-2}$ is applied along -$x$ direction for 0 ns < $t$ < 0.81 ns followed by a relaxation without applying any current until $t$ = 1 ns. At $t$ = 1 ns, a stable skyrmion is in the output side, which represents output = 1.

**Skyrmion logic gates.** Finally we propose logic gates such as the OR and AND gates based on skyrmions. The setup of the OR gate is shown in Fig. 1d, while that of the AND gate is shown in Fig. 1e. First we demonstrate the OR gate in Fig. 5 (see Supplementary Movies S9-S11). The OR gate is an operation such that 0 + 0 = 0, 0 + 1 = 1, 1 + 0 = 1 and 1 + 1 = 1. In skyrmionic logic, binary 0 corresponds to the absence of a skyrmion and binary 1 corresponds to the presence of a skyrmion. The process of 0 + 0 = 0 is trivial, which means that when there is no input, there is no output. We interpret the process 1 + 0 = 1 as follows. When a skyrmion exists in the left input A branch and there is no skyrmion in the left input B branch, one skyrmion is ejected to the right output nanowire. In the same way, the process 0 + 1 = 1 can be interpreted that, when a skyrmion exist in the left input B branch and there is no skyrmion in the left input A branch, one skyrmion is ejected to the right output nanowire. The important process is that 1 + 1 = 1. This is the process that there is a skyrmion both in the left input A and the left input B branches at the first stage. When we apply current, only one skyrmion is ejected. This can be done by merging process as we have already shown in Fig. 4b.

The AND gate (0 + 0 = 0, 0 + 1 = 0, 1 + 0 = 0 and 1 + 1 = 1) can also be realized by using skyrmions. The process 0 + 0 = 0 and 1 + 1 = 1 are the same as that of the OR gate. For implementing the process 1 + 0 = 0 or 0 + 1 = 0, a skyrmion must disappear and there should be no output when a skyrmion exists only at one of the input branch. This is indeed possible by using the sample shown in Fig. 1d. As shown in Fig. 6 (see Supplementary Movies S12-S14), a skyrmion in the left input A branch is converted into a meron in the central region, and disappears by touching a sample edge. The same situation occurs when there is only one skyrmion in the left input B branch. On the other hand, there are skyrmions in both A and B branches, these are converted into a domain-wall pair, resulting in the output of one skyrmion.

## Discussion

We have shown that a skyrmion can be converted into an anti-skyrmion or a bimeron (anti-bimeron). The helicity can be reversed in the conversion process. The merging and duplication are also possible by using a Y-junction geometry. The logic gates such as the AND and OR gates can be realized. It should be noted that multiple functional digital information processing circuits can be easily constructed by combining (a) the logical gates we designed and (b) the conversion process of the spin textures with different quantum numbers ($Q_s$, $Q_v$, $Q_h$). In this way a complete logical architecture might be established with new functionalities to outperform the existing spin logic protocols. Thus our results will pave a way to future skyrmionics.

An interesting interpretation of these conversion mechanisms reads as follows. We connect two samples with different material properties by a narrow nanowire channel. The left and right regions are interpreted as different regions which are characterized by different physical constants. Then the vacuum and elementary topological excitations are different, as determined by the physical constants in each region. For example, an elementary topological excitation in the easy-axis region is a skyrmion, while that in the easy-plane region is a bimeron. In a similar way, the Pontryagin number and the helicity is uniquely determined in each region. Our findings are that we can safely convert topological excitations between different regions. In this sense, the narrow region acts as if it were a channel connecting two different regions. When a topological object enters this channel, it loses its topological numbers such as the Pontryagin number and the helicity. When it is ejected into a different region, a new topological number is assigned to adjust to the new region. Nevertheless, the information that a topological object is injected and ejected is conserved. Thus we can convey the information whether a topological object exists or not.

A skyrmion is topologically protected when the sample is infinitely large. On the other hand, such a topological stability is broken when a skyrmion touches an edge. In the narrow nanowire region, the width is smaller than the skyrmion diameter (see Supplementary Information for the case of the channel width larger than the skyrmion size). Thus, the topological stability of the skyrmion is explicitly broken. Then, the topological number can change. In this sense, there is no barrier in the topological number change.

We discuss how our results are robust for perturbations. As shown in the previous study [27], the conversion mechanism between a skyrmion and a domain-wall pair is very robust for any perturbations. For example, the conversion process is robust in a wide range of material parameters such as saturation magnetization, magnetic anisotropy and the DMI. The ratio between wide and narrow nanowires can be changed as long as the width of the narrow nanowire is less than the skyrmion diameter and that of the wide nanowire is larger than the skyrmion diameter. The conversion mechanism is also robust for interface roughness, the shape of edges, magnetic impurities and thermal fluctuation, as shown in Ref. 27. Furthermore, the exact matching of the Gilbert damping and the non-adiabatic spin-transfer torque coefficient is not necessary since the length scale of our sample is small compared to the skyrmion Hall effect. Accordingly our results are also very robust since the elementary process is based on the conversion mechanism.

## Methods

**The skyrmion number, the vorticity and the helicity.** We parameterize the spin field **n** in the polar coordinate as

$$n_x = \sin[f(r)]\cos(\phi + Q_h), \quad n_y = Q_v \sin[f(r)]\sin(\phi + Q_h), \quad n_z = \cos[f(r)], \tag{1}$$

where $Q_v = \pm 1$ denotes the vorticity of $(n_x, n_y)$ and $Q_h$ denotes the helicity. $x$, $y$, $z$ are the orthogonal coordinate, while $r$ is the radius vector, $\phi$ is the azimuthal angle of the polar cordinate. $f(r)$ is the radius function which determines the $n_z$ configuration. The Pontryagin number is determined as

$$Q_s = -\frac{1}{8\pi}\int d^2x \sum_{ij}\varepsilon_{ij}\mathbf{n}(\mathbf{x})\cdot\left(\partial_i\mathbf{n}(\mathbf{x})\times\partial_j\mathbf{n}(\mathbf{x})\right) = \frac{Q_v}{2}\left[\lim_{r\to\infty}\cos f(r) - \cos f(0)\right], \tag{2}$$

which is determined by the product of the vorticity and the difference between the spin direction of the core and the tail of the skyrmion. We show typical examples of skyrmions, merons and bimerons in Supplementary Figure 1.

**Hamiltonian and energy.** Our system is the 2-dimensional magnet. The Hamiltonian reads

$$H = H_A + H_K + H_{DM} + H_Z. \tag{3}$$

Each term reads as follows: $H_A$ describes the nonlinear O(3) sigma model,

$$H_A = A\int d^2x(\partial_k \mathbf{n})^2, \tag{4}$$

$H_K$ the easy-axis anisotropy,

$$H_K = -K\int d^2x(n_z)^2, \tag{5}$$

$H_{DM}$ the DMI,

$$H_{DM} = \int d^2x D_\perp \left[n_z\left(\frac{\partial n_x}{\partial x} + \frac{\partial n_y}{\partial y}\right) - n_x\frac{\partial n_z}{\partial x} - n_y\frac{\partial n_z}{\partial y}\right]$$

$$+ D_\parallel \left[n_z\left(\frac{\partial n_x}{\partial y} - \frac{\partial n_y}{\partial x}\right) - n_x\frac{\partial n_z}{\partial y} + n_y\frac{\partial n_z}{\partial x}\right]$$

$$= \int d^2x\, D_\perp[n_z\,\mathrm{div}\,\mathbf{n} - (\mathbf{n}\cdot\nabla)n_z] - D_\parallel \mathbf{n}(\mathbf{x})\cdot(\nabla\times\mathbf{n}(\mathbf{x})), \tag{6}$$

where $D_\perp(D_\parallel)$ is the Néel-type (Bloch-type) DMI and $H_Z$ is the Zéeman effect,

$$H_Z = -B\int d^2x\, n_z(\mathbf{x}), \tag{7}$$

with $B > 0$. Here, $J$ is the exchange energy, $K$ is the single-ion easy-axis-anisotropy constant, and $\mathbf{n} = (n_x,$

$n_y$, $n_z$) = **M**/$M_S$ is a classical spin field of unit length.

By substituting (1), we obtain Hamiltonian

$$H = \int 2\pi r dr \left[ A \left[ (\partial_r f(r))^2 + \left( \frac{1}{r^2} + \frac{2K}{A} \right) \sin^2[f(r)] \right] \right.$$
$$\left. + D \left[ \partial_r f(r) + \frac{\sin[2f(r)]}{2r} \right] - B \cos f(r) \right], \quad (8)$$

with $D = D_\perp \cos Q_h + D_\parallel \sin Q_h$ for $Q_v = 1$, $D = 0$ for $Q_v = -1$. We obtain the Néel-type Skyrmion ($Q_h = 0$, π) for $D_\parallel = 0$, while the Bloch-type Skyrmion ($Q_h = \pi/2$, $3\pi/2$) for $D_\perp = 0$. The DM interaction plays a crucial role for the stability of topological structures such as a skyrmion and meron. For $Q_v = -1$, the DMI does not contribute to the stability. Thus the topological structure with $Q_v = -1$ is not dynamically stable due to the Derrick-Hobart theorem [36, 37]

The strong localization of its core allows us to use a linear ansatz [38, 39]. We set $\kappa = \lim_{r \to \infty} \cos f(r)$. We define

$$f(r) = \begin{cases} \pi(1 - r/R) & \text{for} \quad 0 \leq r \leq R \\ 0 & \text{for} \quad r \geq R \end{cases} \quad (9)$$

in the case of $\kappa = 1$, while we define

$$f(r) = \begin{cases} \pi(r/R) & \text{for} \quad 0 \leq r \leq R \\ \pi & \text{for} \quad r \geq R \end{cases} \quad (10)$$

in the case $\kappa = -1$. By substituting them into equation (7), we obtain

$$H = K\pi R^2 + A\pi(\gamma - \text{Ci}(2\pi) + \log 2\pi) - \kappa D R \pi^2 + \kappa \frac{4}{\pi} B R^2$$
$$= \left( K\pi + \kappa \frac{4}{\pi} B \right)(R - R_{Sk})^2 - \frac{D^2 \pi^4}{4K\pi + \frac{16}{\pi}\kappa B} + A\pi(\pi^2 + \gamma - \text{Ci}(2\pi) + \log 2\pi), \quad (11)$$

where $\gamma = -0.58$ is the Euler's gamma constant and Ci is the cosine integral defined by $\text{Ci}(x) = -\int_x^\infty \frac{\cos t}{t} dt$ with $\text{Ci}(2\pi) = -0.023$. The skyrmion radius is determined as

$$R_{Sk} = \frac{\kappa D \pi^2}{2K\pi + \frac{8}{\pi}\kappa B}, \quad (12)$$

with the energy

$$E_{Sk} = -\frac{D^2 \pi^4}{4K\pi + \frac{16}{\pi}\kappa B} + 38.7A. \quad (13)$$

We consider the case $D_\parallel = 0$. We note that the radius must be positive, which leads $\kappa D > 0$ for $K > 0$. The minimal energy occurs when $Q_h = 0$ for $\kappa = 1$, while it occurs at $Q_h = \pi$ for $\kappa = -1$. Hence the helicity is reversed when a skyrmion is converted into an anti-skyrmion. See Fig. 2c and Supplementary Movie S3.

**Modeling and simulation.** The micromagnetic simulations are performed using the Object Oriented MicroMagnetic Framework (OOMMF) including the Dzyaloshinskii-Moriya interaction (DMI) extended module [29-32]. The time-dependent magnetization dynamics is governed by the Landau-Lifshitz-Gilbert (LLG) equation including spin torque [40-44]. The average energy density $E$ is a function of **M**, which contains the exchange energy term, the anisotropy energy term, the applied field (Zeeman) energy term, the magnetostatic (demagnetization) energy term and the DMI energy term.

For micromagnetic simulations, we consider 1-nm-thick cobalt nanotracks with length of 450 ~ 600 nm and width of 10 ~ 150 nm on the substrate. The intrinsic magnetic parameters are adopted from Refs. 5 & 29: Gilbert damping coefficient $\alpha = 0.3$ and the value for Gilbert gyromagnetic ratio is $-2.211 \times 10^5$ m A$^{-1}$ s$^{-1}$. Saturation magnetization $M_S = 580$ kA m$^{-1}$, exchange stiffness $A = 15$ pJ m$^{-1}$, DMI constant $D = 3.5$ mJ m$^{-2}$ and perpendicular magnetic anisotropy (PMA) $K = 0.8$ MJ m$^{-3}$ unless otherwise specified. Thus, the exchange length is $l_{ex} = \sqrt{\frac{A}{K}} = 4.3$ nm. All samples are discretized into cells of 1 nm × 1 nm × 1 nm in the simulation, which is sufficiently smaller than the exchange length and the skyrmion size to ensure the numerical accuracy.

For all simulation reported throughout this paper, the skyrmion is firstly created at the center of the left input side of the nanowire by a spin current perpendicular to the plane (CPP). Then the system is relaxed to an energy minimum state without applying any current. Next, we start the timer and the spin current ($P = 1$) is injected into the nanowire with the geometry of current-in-plane (CIP). In the default configuration, the electrons flow toward the right, *i.e.*, the current flows toward the left. As shown in Refs. 27 & 29, in order to ensure the skyrmion moves along the central line of the nanowire without additional transverse motion, the non-adiabatic torque coefficient is also set to the same value of the damping coefficient in our work, *i.e.*, $\beta = \alpha = 0.3$.

## Acknowledgments

Y.Z. thanks the support by the Seed Funding Program for Basic Research and Seed Funding Program for Applied Research from the University of Hong Kong, ITF Tier 3 funding (ITS/171/13) and University Grants Committee of Hong Kong (Contract No. AoE/P-04/08). M.E. thanks the support by the Grants-in-Aid for Scientific Research from the Ministry of Education, Science, Sports and Culture, No. 25400317. X.C.Z. thanks M.J. Donahue for useful discussion. M.E. is very much grateful to N. Nagaosa for many helpful discussions on the subject.

## Author contributions

X.C.Z. carried out the numerical simulations. M.E. performed the theoretical analysis. Y.Z. coordinated the project. All authors designed the skyrmion logic gates, interpreted the data and contributed to preparing the manuscript and supplementary information. Correspondence and requests for materials should be addressed to M.E. or Y.Z.

## Competing financial interests

The authors declare no competing financial interests.

## How to cite this article

Zhang, X. C., Ezawa, M. & Zhou, Y. Magnetic skyrmion logic gates: conversion, duplication and merging of skyrmions. *Sci. Rep.* **5**, 9400; DOI: 10.1038/srep09400 (2015).